\documentclass[aps,prd,preprint,tightenlines,groupedaddress,showpacs]{revtex4}
\usepackage{epsf,epsfig,graphics,graphicx,amssymb,amsthm}
\begin{document}

\title{Scalar field fluctuations in Schwarzschild-de Sitter space-time}

\author{Hing-Tong Cho$^1$}
\author{Kin-Wang Ng$^2$}
\author{I-Chin Wang$^3$}
\affiliation{
$^1$Department of Physics, Tamkang University, Tamsui, Taiwan 251, R.O.C.\\
$^2$Institute of Physics, Academia Sinica, Taipei, Taiwan 115, R.O.C.\\
$^3$Department of Physics, National Taiwan Normal University,
Taipei, Taiwan 116, R.O.C.}

\date{\today}
%{\today}

\begin{abstract}
We calculate quantum fluctuations of a free scalar field in the
Schwarzschild-de Sitter space-time, adopting the planar coordinates
that is pertinent to the presence of a black hole in an
inflationary universe. In a perturbation approach, doing expansion
in powers of a small black hole event horizon compared to the de Sitter cosmological
horizon, we obtain time evolution of
the quantum fluctuations and then derive the scalar power
spectrum.
\end{abstract}

\maketitle

\section{Introduction}

The cosmic microwave background that we observe today is almost
homogenous and isotropic. The background temperature in our sky is
about $2.7$K with a tiny fluctuation at a level of about
$10^{-5}$K. This is consistent with measurements of the matter
content of the Universe, altogether prevailing a spatially flat
universe. Inflation scenario can explain the homogeneity, the
isotropy, and the flatness of the present Universe~\cite{olive}.
Moreover, quantum fluctuations of the inflaton field during inflation can
give rise to primordial density fluctuations with a nearly
scale-invariant power spectrum which is consistent with the recent
WMAP data on cosmic microwave background
anisotropies~\cite{wmap5a}. Therefore, a general assumption usually
made in most cosmological models is that the background metric is
homogenous and isotropic. For example, a de Sitter metric is used
in the inflationary era and a flat Friedmann-Robertson-Walker
metric is used in the subsequent hot big bang. An interesting
notion is the recent discovery of a dominant component in the
matter content, dubbed the dark energy, which exerts a negative
pressure to drive an accelerated expansion of the Universe~\cite{linder}.
If dark energy is a form of vacuum energy, our Universe will coast to
the de Sitter space-time or the inflating phase again in the
future.

Apparently our present Universe is not so homogeneous and
isotropic because we observe local non-linear structures such as
stars, galaxies, clusters of galaxies, and very massive black
holes. An appropriate space-time, for example, for a massive black
hole sitting in the accelerating Universe, would be described by
the Schwarzschild metric in the vicinity of the black hole and by
the de Sitter metric at places far from the black hole. Presently,
these local structures are decoupled from the Hubble flow, so it
suffices to use the Friedmann-Robertson-Walker metric to study the
large scale structures of the Universe. However, in the early
Universe the gravitational effect of a black hole to the
background metric may be important and should be addressed. For
example, the existence of a black hole or a distribution of black
holes at the onset of inflation can invalidate the use of a
homogenous and isotropic background metric for the calculation of
de Sitter quantum fluctuations. This also applies to the situation
when we are very near to one of these black holes that still
exists today or not.

The cosmic no hair conjecture infers that the inflationary
universe approaches asymptotically the de Sitter space-time till
the end of inflation~\cite{hawking77}. Nevertheless, the effects of
matter and space-time inhomogeneities to inflation should be
considered. Several authors have studied the onset of inflation
under inhomogeneous initial conditions to determine whether large
inhomogeneity during the very early Universe can prevent the
Universe from entering an inflationary era~\cite{inhom}. It was
found that in some cases a large initial inhomogeneity may
suppress the onset of inflation~\cite{gold}. If the inflaton field
is sufficiently inhomogeneous, the wormhole can form from collapsing
vacuum energy density peaks before the inhomogeneity is damped by the exponential
expansion~\cite{holcomb}. In the case of inhomogeneities in a dust
era before inflation, some inhomogeneities can collapse into a
black-hole space-time~\cite{garfinkle}. Furthermore, for the
inhomogeneities of the space-time itself, energies in the form of
gravitational waves can also form a black-hole
space-time~\cite{nakao}. As a consequence, at the onset of
inflation, the distortion of the metric by these inhomogeneities
should be taken into account.

With these considerations in mind, in this work we will
investigate the quantum fluctuations of a free massless scalar field in the
Schwarzschild-de Sitter (SdS) space-time. In the static coordinate
system, the line element of the SdS space-time is given by
\begin{equation}
ds^{2}=-\left(1-{2GM\over r}-H^2r^2\right)dt^{2}+\left(1-{2GM\over
r}-H^2r^2\right)^{-1}dr^{2}+r^{2}d\Omega^{2}, \label{static}
\end{equation}
where $G=M_{\rm Pl}^{-2}$, $M$ is the mass of the black hole, and $H$ is the Hubble
parameter for inflation. Here we use the convention with
$c=\hbar=1$. As is well-known, the SdS metric has a black hole
horizon and a cosmological horizon.
The casual structure of the SdS space-time is depicted in the
Penrose diagram given in, for example, Ref.~\cite{shiromizu}.
In the static coordinates~(\ref{static}) an observer can only receive a signal
inside or just right on the cosmological horizon. This static
metric is insufficient for our purpose because in the cosmological
setting we aim at studying the temporal evolution of a Fourier
mode of the scalar quantum fluctuations that crosses the
cosmological horizon during inflation. Therefore, we will instead
use the planar coordinates for the SdS metric~\cite{shiromizu},
which is given by
\begin{equation}
ds^{2}=-f(r,\tau)d\tau^{2}+h(r,\tau)(dr^{2}+r^{2}d\Omega^{2}),
\label{planar}
\end{equation}
where $d\tau=a^{-1}(\tau)dt$ is the conformal time and
$a(t)=e^{Ht}$. In Eq.~(\ref{planar}), for simplicity we have used the same notations,
$t$ and $r$, actually referring to different local coordinates
than those in Eq.~(\ref{static}).
The $f$ and $h$ functions are given by
\begin{equation}
f(r,\tau)=a^{2}(\tau)\left[1-\frac{GM}{2a(\tau)r}\right]^2
\left[1+\frac{GM}{2a(\tau)r}\right]^{-2},\quad
h(r,\tau)=a^{2}(\tau)\left[1+\frac{GM}{2a(\tau)r}\right]^{4},
\end{equation}
with the cosmic scale factor $a(\tau)=-1/(H\tau)$.
In these coordinates, the black hole horizon corresponds to
$r=GM/(2a)$ and the cosmological horizon is given by $r=a/H$.
For our purpose, we will restrict the range of validity of $\tau$ and $r$
to $-1/H<\tau<0$ and $GM/(2a)<r$.
Note that at late times (i.e., $\tau\rightarrow 0^{-}$) the planar coordinates
behave like a de Sitter expansion.

It is well-known that a black hole evaporates into Hawking
radiation which leads to a mass loss of the
black hole and to its eventual disappearance~\cite{hawking75}.
For a Schwarzschild black hole with mass $M$,
the evaporation time is given by
\begin{equation}
t_{\rm ev}=5120\pi G^2 M^3.
\end{equation}
In general the evaporation time scale of a SdS black hole is
different from a Schwarzschild one. However, for small SdS black holes
with the black hole temperature much higher than the de Sitter
temperature, the evaporation time scale should be of the same order as the
Schwarzschild case.
Therefore, our present consideration requires the condition that
the evaporation time scale of the black hole is
longer than the time scale of inflation, i.e., $H t_{\rm ev}>1$.
This gives the lower bound on the mass of the black hole for
a given inflation scale:
\begin{equation}
\frac{M}{M_{\rm Pl}}> 3.96\times 10^{-2}
\left(\frac{M_{\rm Pl}}{H}\right)^{1\over3}.
\label{ev}
\end{equation}

In the next section, we will review the scalar quantum
fluctuations in the de Sitter metric. Then, we will introduce
a perturbation method to expand the planar metric in powers of the black
hole mass $M$ to find an approximate solution of the scalar equation.
Section~\ref{numerical} contains the numerical
results of the first-order solutions. Section~\ref{conclusion} is
our conclusion.

\section{Perturbation approach to the Schwarzschild-de Sitter space-time}
\label{model}

\subsection{Classical solution}

Consider a massless scalar field $\phi$ which satisfies the Klein-Gordon
equation in the SdS space-time,
\begin{equation}
\partial_{\mu}\left(\sqrt{-g}g^{\mu\nu}\partial_{\nu}\phi\right)=0.
\label{kgeq}
\end{equation}
Since the space is spherically symmetric, one can expand $\phi$ as
\begin{equation}
\phi(x)=\sum_{lm}\varphi_{l}(r,\tau)Y_{lm}(\theta,\phi).
\label{sphphi}
\end{equation}
If one further writes $\varphi_{l}(r,\tau)$ in a spectral form
in terms of spherical Bessel functions $j_{l}$~\cite{Grad},
\begin{equation}
\varphi_{l}(r,\tau)=\int_{0}^{\infty}dk\ \! k^{2}j_{l}(kr)
\varphi_{kl}(\tau),
\end{equation}
then we will have
\begin{equation}
\phi(x)=\int_{0}^{\infty}dk\sum_{lm}\varphi_{klm}(x),\quad
\varphi_{klm}(x)=k^{2}j_{l}(kr)\varphi_{kl}(\tau)Y_{lm}(\theta,\phi).
\end{equation}
We define the spectral function of the fluctuations of the field $\phi(x)$ as
\begin{equation}
S_{kl}(r,\tau)\equiv (2l+1)j_{l}^{2}(kr) P_{kl}(\tau),
\quad P_{kl}(\tau)=\frac{k^5}{4\pi}|\varphi_{kl}(\tau)|^{2}.
\label{spectralkl}
\end{equation}
The power spectrum $P_{kl}$ which gives the power of the fluctuations in a logarithmic
interval of $k$ is useful for comparing theoretical predictions with observations.

Our next task is to calculate the function $\varphi_{kl}(\tau)$ in
Eq.~(\ref{spectralkl}) in the SdS planar metric~(\ref{planar}).
The Klein-Gordon equation~(\ref{kgeq}) beomes
\begin{equation}
-\frac{1}{\sqrt{fh}}\partial_{\tau}\left(\sqrt{\frac{h^{3}}{f}}
\partial_{\tau}\varphi_{l}\right)+\frac{1}{r^{2}\sqrt{fh}}\partial_{r}
\left(r^{2}\sqrt{fh}\partial_{r}\varphi_{l}\right)-\frac{l(l+1)}{r^{2}}\varphi_{l}=0,\label{KGeqn}
\end{equation}
where we have separated out the angular part of $\phi(x)$ in
Eq.~(\ref{sphphi}). There is no exact solution to this equation.
We therefore adopt a perturbative approach assuming that the
quantity,
\begin{equation}
\epsilon\equiv GMH, \label{epsilon}
\end{equation}
is a small parameter. But the condition~(\ref{ev}) implies that
\begin{equation}
\epsilon >  3.96\times 10^{-2}
\left(\frac{H}{M_{\rm Pl}}\right)^{2\over3}.
\end{equation}
However, this shows that the smallness of $\epsilon$
can be easily satisfied for any reasonable inflation scale.
Let us rewrite Eq.~(\ref{KGeqn}) as
\begin{eqnarray}
&&\partial_{\tau}^{2}\varphi_{l}-\frac{2}{\tau}\partial_{\tau}\varphi_{l}-\partial_{r}^{2}
\varphi_{l}-\frac{2}{r}\partial_{r}\varphi_{l}+\frac{l(l+1)}{r^2}\varphi_{l}
=\left(1-\frac{h}{f}\right)\partial_{\tau}^{2}\varphi_{l} -
\frac{2}{\tau}\left(1-\frac{h}{f}\right)\partial_{\tau}\varphi_{l}
\nonumber \\ && -
\frac{1}{\tau^2\sqrt{fh}}\partial_{\tau}\left(\tau^2\sqrt{\frac{h^{3}}{f}}\right)
\partial_{\tau}\varphi_{l} + \frac{1}{\sqrt{fh}}\partial_{r}
\left(\sqrt{fh}\right)\partial_{r}\varphi_{l} \label{KGeqn2}
\end{eqnarray}
We then expand the functions $f$ and $h$ in powers of $\epsilon$
as defined in Eq.~(\ref{epsilon}) and write
\begin{equation}
\varphi_{l}=\varphi_{l}^{(0)}+\varphi_{l}^{(1)}+\varphi_{l}^{(2)}+\cdots
\label{phiexpand}
\end{equation}
as an expansion in orders of $\epsilon$. It is straightforward to
show that the right-hand side of Eq.~(\ref{KGeqn2}) becomes
\begin{equation}
8\left(\frac{\epsilon\tau}{2r}\right)\left(\partial^{2}_{\tau}
\varphi_{l}-\frac{1}{\tau}\partial_{\tau} \varphi_{l}\right) -
30\left(\frac{\epsilon\tau}{2r}\right)^2\left(\partial^{2}_{\tau}
\varphi_{l}-\frac{1}{15\tau}\partial_{\tau} \varphi_{l}
-\frac{1}{15r}\partial_{r} \varphi_{l}\right) + \cdots
\label{fhexpand}
\end{equation}
Note that this is essentially expanded in terms of $\epsilon\tau/(2r)=GM/(2ar)$.
As we are adopting a perturbative approach,
we cannot really take $ar$ close to the black hole horizon $2GM$ and
should put a lower cutoff on $r$ of the order of $2GM/a$ in the calculation.
However, we find that taking the cutoff to zero will not affect the results
that we will obtain below.

\subsubsection{Zeroth order}

From Eqs.~(\ref{KGeqn2}),~(\ref{phiexpand}), and~(\ref{fhexpand}),
the zeroth order corresponds to the de Sitter case with
\begin{equation}
\partial_{\tau}^{2} \varphi_{l} ^{(0)}-\frac{2}{\tau}\partial_{\tau}\varphi_{l} ^{(0)}-\partial_{r}^{2}
\varphi_{l} ^{(0)}-\frac{2}{r}\partial_{r}\varphi_{l}
^{(0)}+\frac{l(l+1)}{r^2}\varphi_{l} ^{(0)}=0. \label{order0}
\end{equation}
To solve this equation, we take the Bessel transform,
\begin{equation}
\varphi_{l} ^{(0)} (r,\tau)=\int_{0} ^{\infty}dkk^{2}j_{l}
(kr)\varphi_{kl} ^{(0)} (\tau)\;\;.
\end{equation}
Then we have
\begin{equation}
\partial_{\tau}^{2} \varphi_{kl} ^{(0)} (\tau)-\frac{2}{\tau}\partial_{\tau}\varphi_{kl}
^{(0)} (\tau)+k^2\varphi_{kl} ^{(0)} (\tau)=0\;\;,
\end{equation}
and the solution is
\begin{equation}
\varphi_{kl} ^{(0)}
(\tau)=C_{1}(-k\tau)^{\frac{3}{2}}H_{\frac{3}{2}} ^{(1)}
(-k\tau)+C_{2}(-k\tau)^{\frac{3}{2}}H_{\frac{3}{2}} ^{(2)}
(-k\tau)\;,
\label{phikl0c1c2}
\end{equation}
where $H_{\frac{3}{2}} ^{(1)}$ and $H_{\frac{3}{2}} ^{(2)}$
are the Hankel functions of order $3/2$~\cite{Grad}.
If we take the boundary conditions:
\begin{equation}
C_{1}=-\frac{H}{k^2\sqrt{2k}},\quad{\rm and}\quad C_{2}=0,
\label{bcon}
\end{equation}
then we will have
\begin{equation}
\varphi_{kl}^{(0)}(\tau)=-\frac{H\tau}{k\sqrt{\pi k}}
\left(1-\frac{i}{k\tau}\right)e^{-ik\tau}, \label{phikl0}
\end{equation}
and
\begin{equation}
P_{kl}^{(0)}(\tau)=\frac{k^5}{4\pi}
\left\vert\varphi_{kl}^{(0)}(\tau)\right\vert^{2}
=\frac{H^2}{4\pi^2}(1+k^2\tau^2).
\end{equation}
As $\tau\rightarrow 0$, $P_{kl}^{(0)}(\tau)\rightarrow
H^2/(4\pi^2)$. This result gives rise to a scale-invariant power
spectrum that is preferred by observational data~\cite{wmap5a}.
Also, it matches the well-known scale-invariant power
spectrum of de Sitter quantum fluctuations~\cite{desitter},
which presumably undergo decoherence to become classical fluctuations.

\subsubsection{First order}

With the perturbative expansion for $\varphi_{l}$ in
Eq.~(\ref{phiexpand}), the Klein-Gordon equation in
Eq.~(\ref{KGeqn2}) can be solved perturbatively. The first order
$\varphi_{l}^{(1)}(r,\tau)$ then satisfies
\begin{equation}
\partial^{2}_{\tau} \varphi^{(1)}_{l}-\frac{2}{\tau}\partial_{\tau}
\varphi^{(1)}_{l}-\partial_{r} ^{2}
\varphi^{(1)}_{l}-\frac{2}{r}\partial_{r}\varphi^{(1)}_{l}+\frac{l(l+1)}{r^{2}}\varphi^{(1)}_{l}=J_{1}
\;,\label{1oe}
\end{equation}
where the source term is given by
\begin{eqnarray}
J_{1}(r,\tau)&=&\frac{4\epsilon\tau}{r}\left(\partial^{2}_{\tau}
\varphi^{(0)}_{l}-\frac{1}{\tau}\partial_{\tau} \varphi^
{(0)}_{l}\right)\nonumber\\
&=&\frac{4\epsilon H\tau^{2}}{\sqrt{\pi}r}
\int^{\infty}_{0}dkk^{2}j_{l}(kr)(k^{\frac{1}{2}}e^{-ik\tau}).
\label{1os}
\end{eqnarray}
To solve this inhomogeneous equation, we use the Green's function
$G(r,\tau;r',\tau')$ which satisfies the equation,
\begin{equation}
\partial^{2}_{\tau} G-\frac{2}{\tau}\partial_{\tau}
G-\partial_{r} ^{2}
G-\frac{2}{r}\partial_{r}G+\frac{l(l+1)}{r^{2}}G=\frac{\delta(r-r')\delta(\tau-\tau')}{r^{2}}
\;.\label{Geqn}
\end{equation}
Using the completeness property of the spherical Bessel functions,
\begin{equation}
\int_{0} ^{\infty}dkk^{2}\left[\sqrt{\frac{2}{\pi}}rj_{l}
(kr)\right]\left[\sqrt{\frac{2}{\pi}}r'j_{l}
(kr')\right]=\delta(r-r'),
\end{equation}
and taking
\begin{equation}
G_{l}(r,\tau;r',\tau')=\int^{\infty}_{0}dkk^{2}g_{k}(\tau,\tau')j_{l}(kr)j_{l}(kr')\;,
\end{equation}
Eq.~(\ref{Geqn}) becomes
\begin{equation}
\partial^{2}_{\tau}g_{k}-\frac{2}{\tau}\partial_{\tau}g_{k}+k^{2}g_{k}=\frac{2}{\pi}\delta(\tau-\tau')\;.
\end{equation}
For the retarded Green's function, $g_{k}=0$ for
$\tau'>\tau>\tau_{i}$, where $\tau_i$ denotes an initial time when
the source begins to operate. For $0>\tau>\tau'$,
\begin{eqnarray}
g_{k}(\tau,\tau')&=&\frac{i}{2\tau'^{2}k^{3}}\left[(-k\tau)^{\frac{3}{2}}
H^{(1)}_{\frac{3}{2}}(-k\tau)
(-k\tau')^{\frac{3}{2}}H^{(2)}_{\frac{3}{2}}(-k\tau')\right.\nonumber\\
&&\ \ \ \ \ \ \ \ \ \
\left.-(-k\tau')^{\frac{3}{2}}H^{(1)}_{\frac{3}{2}}(-k\tau')
(-k\tau)^{\frac{3}{2}}H^{(2)}_{\frac{3}{2}}(-k\tau)\right].
\end{eqnarray}
With this retarded Green's function, the first order
$\varphi_{l}^{(1)}(r,\tau)$ can be expressed as
\begin{equation}
\varphi^{(1)}_{l}(r,\tau)= \int^{\infty}_{0}dkk^{2}j_{l}(kr)
\varphi_{kl}^{(1)}(\tau) = \int^{\infty}_{0}dr'r'^{2}
\int^{0}_{\tau_{i}}d\tau' G(r,\tau;r',\tau')J_{1}(r',\tau').
\end{equation}
Hence, we find that
\begin{eqnarray}
\varphi_{kl}^{(1)}(\tau)&=&\frac{2i\epsilon H}{\sqrt{\pi}k^{3}}
\int^{\infty}_{0}dk'k'^{2}(k'^{\frac{1}{2}})\int^{\infty}_{0}dr'r'
j_{l}(kr')j_{l}(k'r') \int^{\tau}_{\tau_{i}}d\tau'
e^{-ik'\tau'}\times \nonumber\\ &&
\left[(-k\tau)^{\frac{3}{2}}H^{(1)}_{\frac{3}{2}}(-k\tau)
(-k\tau')^{\frac{3}{2}}H^{(2)}_{\frac{3}{2}}(-k\tau')-(-k\tau')^{\frac{3}{2}}H^{(1)}_{\frac{3}{2}}(-k\tau')
(-k\tau)^{\frac{3}{2}}H^{(2)}_{\frac{3}{2}}(-k\tau)\right]\;.\nonumber\\
\label{varphiklone}
\end{eqnarray}

It is useful to rewrite $\varphi_{kl}^{(1)}(\tau)$ as
\begin{equation}
\varphi_{kl}^{(1)}(\tau)=
\epsilon\left[\alpha_{kl}(\tau)\varphi_{kl}^{(0)}(\tau)+\beta_{kl}(\tau){\varphi_{kl}^{(0)}}^{*}(\tau)\right].
\end{equation}
The integral over $r'$ can be performed,
\begin{equation}
\int^{\infty}_{0}dr'r'j_{l}(kr')j_{l}(k'r')=\frac{\pi}{2\sqrt{k_{<}k_{>}}}\left(\frac{1}{k_{>}}\right)
\left(\frac{k_{<}}{k_{>}}\right)^{l+\frac{1}{2}}
\frac{\Gamma(l+1)}{\Gamma(l+\frac{3}{2})\Gamma(\frac{1}{2})}F\left(l+1,\frac{1}{2};l+\frac{3}{2},\frac{k_{<}
^{2}}{k_{>}^{2}}\right)\;,
\label{r-integral}
\end{equation}
where $\Gamma$ is the Gamma function, $F$ is the hypergeometric function, and
$k_{<}$ ($k_{>}$) represents the smaller (bigger) one of $k$ and $k'$ ~\cite{Grad}.
After some rescalings the coefficients $\alpha_{kl}$ and
$\beta_{kl}$ can be simplified to
\begin{eqnarray}
\alpha_{kl}(\tau)&=&\frac{-2i\Gamma(l+1)}
{\Gamma(l+\frac{3}{2})\Gamma(\frac{1}{2})} \left[
\int_{0}^{1}dk'k'^{l+\frac{5}{2}}F\left(l+1,\frac{1}{2};l+\frac{3}{2};k'^{2}\right)
\int_{k\tau_{i}}^{k\tau}d\tau' e^{-ik'\tau'}e^{i\tau'}\left(\tau'+i\right) \right. \nonumber \\
&& +\left.
\int_{1}^{\infty}dk'k'^{-l+\frac{1}{2}}F\left(l+1,\frac{1}{2};l+\frac{3}{2};\frac{1}{k'^{2}}\right)
\int_{k\tau_{i}}^{k\tau}d\tau' e^{-ik'\tau'}e^{i\tau'}\left(\tau'+i\right)
\right]
\label{alphakl}
\end{eqnarray}
and
\begin{eqnarray}
\beta_{kl}(\tau)&=&\frac{2i\Gamma(l+1)}
{\Gamma(l+\frac{3}{2})\Gamma(\frac{1}{2})} \left[
\int_{0}^{1}dk'k'^{l+\frac{5}{2}}F\left(l+1,\frac{1}{2};l+\frac{3}{2};k'^{2}\right)
\int_{k\tau_{i}}^{k\tau}d\tau' e^{-ik'\tau'}e^{-i\tau'}\left(\tau'-i\right) \right.\nonumber \\
&& +\left.
\int_{1}^{\infty}dk'k'^{-l+\frac{1}{2}}F\left(l+1,\frac{1}{2};l+\frac{3}{2};\frac{1}{k'^{2}}\right)
\int_{k\tau_{i}}^{k\tau}d\tau' e^{-ik'\tau'}e^{-i\tau'}\left(\tau'-i\right)\right]\;.
\label{betakl}
\end{eqnarray}
The power spectrum in this order is then given by
\begin{equation}
P_{kl}(\tau)=P_{kl}^{(0)}(\tau)\left[1+\epsilon \Delta^{(1)}_{kl}(\tau)\right]\;,
\label{plk}
\end{equation}
where using $\varphi_{kl}^{(1)}$ in Eq.~(\ref{varphiklone}) (noting that it has inside an $\epsilon$),
we have introduced the dimensionless quantity,
\begin{eqnarray}
\Delta^{(1)}_{kl}(\tau) &=& 2\,\epsilon^{-1}\left\vert\varphi^{(0)}_{kl}(\tau)\right\vert^{-2}
{\rm Re}\left[\varphi_{kl}^{(0)}(\tau){\varphi_{kl}^{(1)}}^*(\tau)\right]
\nonumber \\
&=&2\,{\rm Re}\left[\alpha_{kl}(\tau)\right]+2 \left\vert\varphi^{(0)}_{kl}(\tau)\right\vert^{-2}
{\rm Re}\left[\beta_{kl}^{*}(\tau)\varphi_{kl}^{(0)2}(\tau)\right].
\label{deltalk}
\end{eqnarray}
As $\tau\rightarrow 0$, from Eq.~(\ref{phikl0}),
\begin{equation}
\Delta^{(1)}_{kl}(0) = 2\,{\rm Re}
\left[\alpha_{kl}(0)-\beta_{kl}(0)\right].
\label{deltalk0}
\end{equation}
From Eqs.~(\ref{alphakl}) and (\ref{betakl}), we find that
\begin{eqnarray}
\Delta^{(1)}_{kl}(0)&=& \frac{8\Gamma(l+1)}{\Gamma(l+\frac{3}{2})\Gamma(\frac{1}{2})}\left[
\int_{0}^{1}dk'k'^{l+\frac{5}{2}}F\left(l+1,\frac{1}{2};l+\frac{3}{2};k'^{2}\right)\right.+
\nonumber \\
&&\left.\int_{1}^{\infty}dk'k'^{-l+\frac{1}{2}}F\left(l+1,\frac{1}{2};l+\frac{3}{2};\frac{1}{k'^{2}}\right)
\right] \times \nonumber \\
&&\frac{1}{(k'^2-1)^2}\left\{\cos(k\tau_i)\left[2\sin(k'k\tau_i)-k'(k'^2-1)k\tau_i\cos(k'k\tau_i)\right]\right.\nonumber \\
&&\left.+\sin(k\tau_i)\left[k'(k'^2-3)\cos(k'k\tau_i)-(k'^2-1)k\tau_i\sin(k'k\tau_i)\right]\right\}\,.
\end{eqnarray}

\subsection{Quantization}

A unique mode function $\varphi_{klm}(x)$ can be obtained once an appropriate
vacuum is chosen. By using this mode function the scalar field can be
quantized in the standard manner,
\begin{equation}
\hat{\phi}(x)=\int_{0}^{\infty}dk
\sum_{lm}\left[\hat{a}_{klm}\varphi_{klm}(x)
+\hat{a}^{\dagger}_{klm}\varphi_{klm}^{*}(x)\right],
\end{equation}
with the commutation relations:
\begin{eqnarray}
&&[\hat{a}_{klm},\hat{a}_{k'l'm'}]=[\hat{a}^{\dagger}_{klm},\hat{a}^{\dagger}_{k'l'm'}]=0,\nonumber\\
&&[\hat{a}_{klm},\hat{a}^{\dagger}_{k'l'm'}]=\delta(k-k')\delta_{ll'}\delta_{mm'}.
\end{eqnarray}
We have the delta function in $k$ because there is no coupling
between different $k$-modes. We will show this explicitly later by
a perturbation approach. The delta functions in $l$ and $m$ stem
from the rotational invariance about the central black hole. The
vacuum state is defined as
\begin{equation}
\hat{a}_{klm}|0\rangle=0.
\end{equation}

Now $\varphi_{klm}(x)$ is the mode function, so it should also satisfy the Klein-Gordon equation.
Since the space is spherically symmetric, one can write
\begin{equation}
\varphi_{klm}(x)=\varphi_{kl}(r,\tau)Y_{lm}(\theta,\phi).
\end{equation}
But $\varphi_{kl}(r,\tau)$ cannot in general be separated as a product of functions
with only one variable like $k^{2}j_{l}(kr)\varphi_{kl}(\tau)$ as we have done in the classical solution.
It is because $\varphi_{klm}(x)$ is required to satisfy the Klein-Gordon equation
for each $k$, $l$, and $m$, while for the classical wave only $\phi(x)$
itself is required to do so.
However, $\varphi_{kl}(r,\tau)$ can still be obtained perturbatively.
The Klein-Gordon equation for the mode function is the same as Eq.~(\ref{KGeqn}), given by
\begin{equation}
-\frac{1}{\sqrt{fh}}\partial_{\tau}\left(\sqrt{\frac{h^{3}}{f}}
\partial_{\tau}\varphi_{kl}(r,\tau)\right)+\frac{1}{r^{2}\sqrt{fh}}\partial_{r}
\left(r^{2}\sqrt{fh}\partial_{r}\varphi_{kl}(r,\tau)\right)-\frac{l(l+1)}{r^{2}}\varphi_{kl}(r,\tau)=0.
\end{equation}
The two-point correlation function is then given by
\begin{eqnarray}
\langle 0|\hat{\phi}(x)\hat{\phi}(x')|0\rangle&=&
\int_{0}^{\infty}dk\sum_{lm}\varphi_{klm}(x)\varphi^{*}_{klm}(x') \nonumber \\
&=&\int_{0}^{\infty}dk\sum_{l}\frac{2l+1}{4\pi} \varphi_{kl}(r,\tau) \varphi^{*}_{kl}(r',\tau)
 P_l(\cos\gamma),
\end{eqnarray}
where $\gamma$ is the separation angle between the two points. As
$x'\rightarrow x$, we have
\begin{equation}
\langle 0|\hat{\phi}^{2}(x)|0\rangle=\int_{0}^{\infty}\frac{dk}{k}
\sum_{l}\frac{2l+1}{4\pi}k \left|\varphi_{kl}(r,\tau)\right|^{2}.
\end{equation}
In terms of $\varphi_{kl}(r,\tau)$, the spectral function of the fluctuations
of the quantum field $\phi(x)$ can be defined as
\begin{equation}
S_{kl}(r,\tau)\equiv (2l+1)j_{l}^{2}(kr) P_{kl}(r,\tau), \quad
P_{kl}(r,\tau)=\frac{k}{4\pi j_{l}^{2}(kr)}\left|\varphi_{kl}(r,\tau)\right|^{2}.
\end{equation}
Perturbatively,
\begin{eqnarray}
P_{kl}(r,\tau)&=&\frac{k}{4\pi j_{l}^{2}(kr)}\left|\varphi_{kl}^{(0)}(r,\tau)+\varphi_{kl}^{(1)}(r,\tau)+\cdots\right|^{2}\nonumber\\
&=&P_{kl}^{(0)}(r,\tau)\left[1+\epsilon \Delta^{(1)}_{kl}(r,\tau)+\cdots\right],
\end{eqnarray}
where we have defined
\begin{eqnarray}
P_{kl}^{(0)}(r,\tau)&=&\frac{k}{4\pi j_{l}^{2}(kr)}\left|\varphi_{kl}^{(0)}(r,\tau)\right|^{2},\label{pklrtau0}\\
\Delta^{(1)}_{kl}(r,\tau)&=&2\,\epsilon^{-1} \left|\varphi_{kl}^{(0)}(r,\tau)\right|^{-2}
{\rm Re}\left[\varphi_{kl}^{(0)}(r,\tau){\varphi_{kl}^{(1)}}^*(r,\tau)\right].
\label{dklrtau1}
\end{eqnarray}

\subsubsection{Zeroth order}

Following the same steps in Eqs.~(\ref{KGeqn2}), (\ref{phiexpand}), and (\ref{fhexpand}),
to the zeroth order we have as before
\begin{equation}
\partial_{\tau}^{2} \varphi_{kl} ^{(0)}(r,\tau)-\frac{2}{\tau}\partial_{\tau}\varphi_{kl} ^{(0)}(r,\tau)-\partial_{r}^{2}
\varphi_{kl} ^{(0)}(r,\tau)-\frac{2}{r}\partial_{r}\varphi_{kl}
^{(0)}(r,\tau)+\frac{l(l+1)}{r^2}\varphi_{kl} ^{(0)}(r,\tau)=0,
\end{equation}
and the general solution is found to be
\begin{equation}
\varphi_{kl}^{(0)}(r,\tau)=k^{2}j_{l}(kr)\varphi_{kl}^{(0)}(\tau),
\label{phiklrtau0}
\end{equation}
where $\varphi_{kl}^{(0)}(\tau)$ is given by Eq.~(\ref{phikl0c1c2}).
The boundary conditions~(\ref{bcon}) indeed correspond to the choice of the Bunch-Davies vacuum
that selects the mode function:
\begin{equation}
\varphi_{kl}^{(0)}(\tau)=-\frac{H\tau}{k\sqrt{\pi k}}
\left(1-\frac{i}{k\tau}\right)e^{-ik\tau},
\end{equation}
and hence the zeroth order power spectrum in Eq.~(\ref{pklrtau0}) is
\begin{equation}
P_{kl}^{(0)}(r,\tau)=\frac{k}{4\pi j_{l}^{2}(kr)}\left|\varphi_{kl}^{(0)}(r,\tau)\right|^{2}
=\frac{H^{2}}{4\pi^{2}}(1+k^{2}\tau^{2}).
\end{equation}

\subsubsection{First order}

To the next order, we have
\begin{equation}
\partial^{2}_{\tau} \varphi^{(1)}_{kl}(r,\tau)-\frac{2}{\tau}\partial_{\tau}
\varphi^{(1)}_{kl}(r,\tau)-\partial_{r} ^{2}
\varphi^{(1)}_{kl}(r,\tau)-\frac{2}{r}\partial_{r}\varphi^{(1)}_{kl}(r,\tau)+\frac{l(l+1)}{r^{2}}\varphi^{(1)}_{kl}(r,\tau)=J_{1}
\;,
\end{equation}
where
\begin{eqnarray}
J_{1}(r,\tau)&=&\frac{4\epsilon\tau}{r}\left[\partial^{2}_{\tau}
\varphi^{(0)}_{kl}(r,\tau)-\frac{1}{\tau}\partial_{\tau} \varphi^
{(0)}_{kl}(r,\tau)\right]\nonumber\\
&=&\frac{4\epsilon H\tau^{2}}{\sqrt{\pi}r}\left[
k^{2}j_{l}(kr)(k^{\frac{1}{2}}e^{-ik\tau})\right].
\end{eqnarray}
The retarded Green's function necessary to solve this equation is the same as before. Then,
\begin{eqnarray}
&&\varphi^{(1)}_{kl}(r,\tau)\nonumber\\
&=& \int^{\infty}_{0}dr'r'^{2}
\int^{0}_{\tau_{i}}d\tau' G(r,\tau;r',\tau')J_{1}(r',\tau')\nonumber\\
&=&\frac{2i\epsilon H\;k^{5\over2}}{\sqrt{\pi}}
\int^{\infty}_{0}\frac{dk'}{k'}j_{l}(k'r)\int^{\infty}_{0}dr'r'
j_{l}(kr')j_{l}(k'r') \int^{\tau}_{\tau_{i}}d\tau'
e^{-ik\tau'}\times \nonumber\\ &&
\left[(-k'\tau)^{\frac{3}{2}}H^{(1)}_{\frac{3}{2}}(-k'\tau)
(-k'\tau')^{\frac{3}{2}}H^{(2)}_{\frac{3}{2}}(-k'\tau')-(-k'\tau')^{\frac{3}{2}}H^{(1)}_{\frac{3}{2}}(-k'\tau')
(-k'\tau)^{\frac{3}{2}}H^{(2)}_{\frac{3}{2}}(-k'\tau)\right].\nonumber\\
\end{eqnarray}
This is in the form of
\begin{equation}
\varphi^{(1)}_{kl}(r,\tau)=\epsilon\int_{0}^{\infty}\frac{dk'}{k'}\left[\alpha_{kk'l}(\tau)\varphi^{(0)}_{k'l}(r,\tau)
+\beta_{kk'l}(\tau){\varphi^{(0)}_{k'l}}^{*}(r,\tau)\right],
\label{phiklrtau1}
\end{equation}
where
\begin{eqnarray}
\alpha_{kk'l}(\tau)&=&-\left(\frac{4i}{\pi}\right)k^{5\over2}{k'}^{1\over2}\int_{0}^{\infty}dr'\;r'j_{l}(kr')j_{l}(k'r')
\int_{\tau_{i}}^{\tau}d\tau'(k'\tau'+i)e^{-i(k-k')\tau'},\label{akk'ltau}\\
\beta_{kk'l}(\tau)&=&\left(\frac{4i}{\pi}\right)k^{5\over2}{k'}^{1\over2}\int_{0}^{\infty}dr'\;r'j_{l}(kr')j_{l}(k'r')
\int_{\tau_{i}}^{\tau}d\tau'(k'\tau'-i)e^{-i(k+k')\tau'}.
\label{bkk'ltau}
\end{eqnarray}
As the perturbative formalism in quantum mechanics, the eigenfunctions in the first order consist of zeroth order eigenfunctions of different energies.

Now we calculate the leading correction~(\ref{dklrtau1}) to the power spectrum.
Substituting Eq.~(\ref{phiklrtau0}) in Eq.~(\ref{phiklrtau1}), we have
\begin{equation}
\Delta^{(1)}_{kl}(r,\tau)=\int_{0}^{\infty}\frac{dk'}{k'} \frac{k'^2}{k^2}
\frac{j_l(k'r)}{j_l(kr)} \left|\varphi_{kl}^{(0)}(\tau)\right|^{-2}
2\,{\rm Re}\left[\alpha^{*}_{kk'l}(\tau) \varphi_{kl}^{(0)}(\tau) \varphi_{k'l}^{(0)*}(\tau)+
\beta^{*}_{kk'l}(\tau) \varphi_{kl}^{(0)}(\tau) \varphi_{k'l}^{(0)}(\tau)\right].
\label{deltalkr}
\end{equation}
As $\tau\rightarrow 0$, this becomes
\begin{equation}
\Delta^{(1)}_{kl}(r,0)=\int_{0}^{\infty}\frac{dk'}{k'} {\left(\frac{k}{k'}\right)}^{1\over2}
\frac{j_l(k'r)}{j_l(kr)}\,2\, {\rm Re}\left[\alpha_{kk'l}(0)-\beta_{kk'l}(0)\right].
\label{delta1klr0}
\end{equation}
Here we obtain from Eqs.~(\ref{akk'ltau}) and (\ref{bkk'ltau}) that
\begin{eqnarray}
&&2\,{\rm Re}\left[\alpha_{kk'l}(0)-\beta_{kk'l}(0)\right]\nonumber\\
&=&\frac{16}{\pi} {\left(\frac{k}{k'}\right)}^{1\over2} k^2 \int_{0}^{\infty}dr'\;r'j_{l}(kr')j_{l}(k'r')\times\nonumber\\
&&\left(\frac{k^2}{k'^2}-1\right)^{-2}\left\{\cos(k'\tau_i)\left[2\sin(k\tau_i)
-\left(\frac{k^2}{k'^2}-1\right)k\tau_i\cos(k\tau_i)\right]\right.\nonumber \\
&&\left.+\sin(k'\tau_i)\left[\frac{k}{k'}\left(\frac{k^2}{k'^2}-3\right)
\cos(k\tau_i)-\left(\frac{k^2}{k'^2}-1\right)k'\tau_i\sin(k\tau_i)\right]\right\}\,,
\end{eqnarray}
where the remaining integral over $r'$ can be evaluated using the formula~(\ref{r-integral}).
This quantity diverges as $k\rightarrow k'$. However, we find that the integral in Eq.~(\ref{delta1klr0})
is finite and weakly dependent on $r$. In the limit of $r\rightarrow 0$, it can be approximated by
\begin{equation}
\Delta^{(1)}_{kl}(r,0)\simeq \int_{0}^{\infty}\frac{dk'}{k'} {\left(\frac{k'}{k}\right)}^{l-{1\over2}}
\,2\, {\rm Re}\left[\alpha_{kk'l}(0)-\beta_{kk'l}(0)\right].
\end{equation}

\section{Numerical Results}
\label{numerical}

Eqs.~(\ref{deltalk}) and (\ref{deltalkr}) are the main results of the present
paper. However, they are complicated integrals and are not
illuminating. Therefore, we perform numerical
calculations of the black-hole corrections to the power spectrum.
Assume that inflation begins
at the initial time $t=0$ and $a(t=0)=1$. Then, the
initial conformal time is $\tau_{i}=-1/H$ and the final conformal
time is $\tau=-e^{-Ht}/H$, where $Ht$ is the e-folding
number of inflation. It is useful to note that the Fourier mode
with momentum $k$ crosses out the cosmological horizon at time $\tau=-1/k$,
or equivalently, when the e-folding number is $Ht=\ln(k/H)$.

\begin{figure}[htp]
\centering
\includegraphics[width=0.5\textwidth]{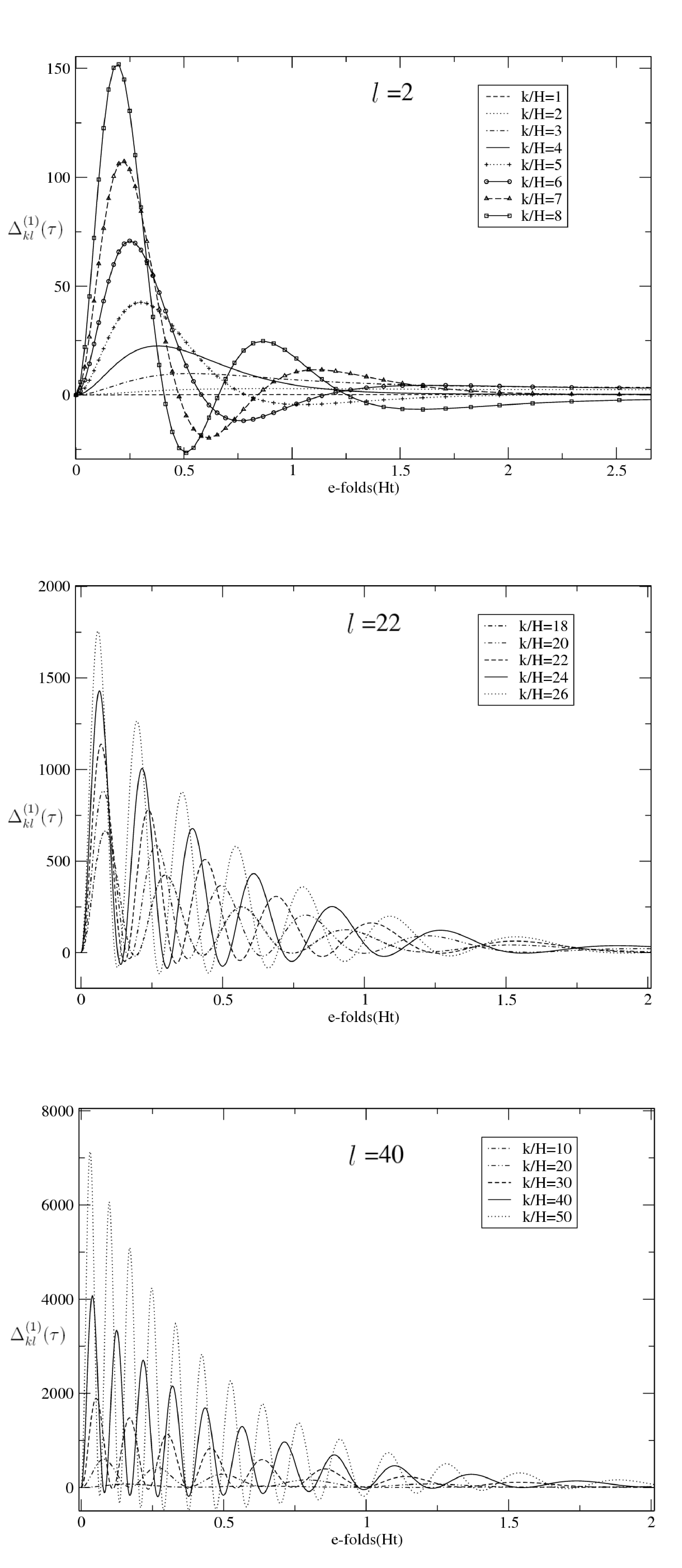}
\caption{Time evolution of the first-order normalized power spectra
$\Delta^{(1)}_{kl}(\tau)$ in Eq.~(\ref{deltalk}) for $l=2,22,40$.
For each case, we have selected some $k$-modes.
The time is expressed in terms of the e-folding number $Ht$.}
\label{fig1}
\end{figure}

\begin{figure}[htp]
\centering
\includegraphics[width=0.68\textwidth]{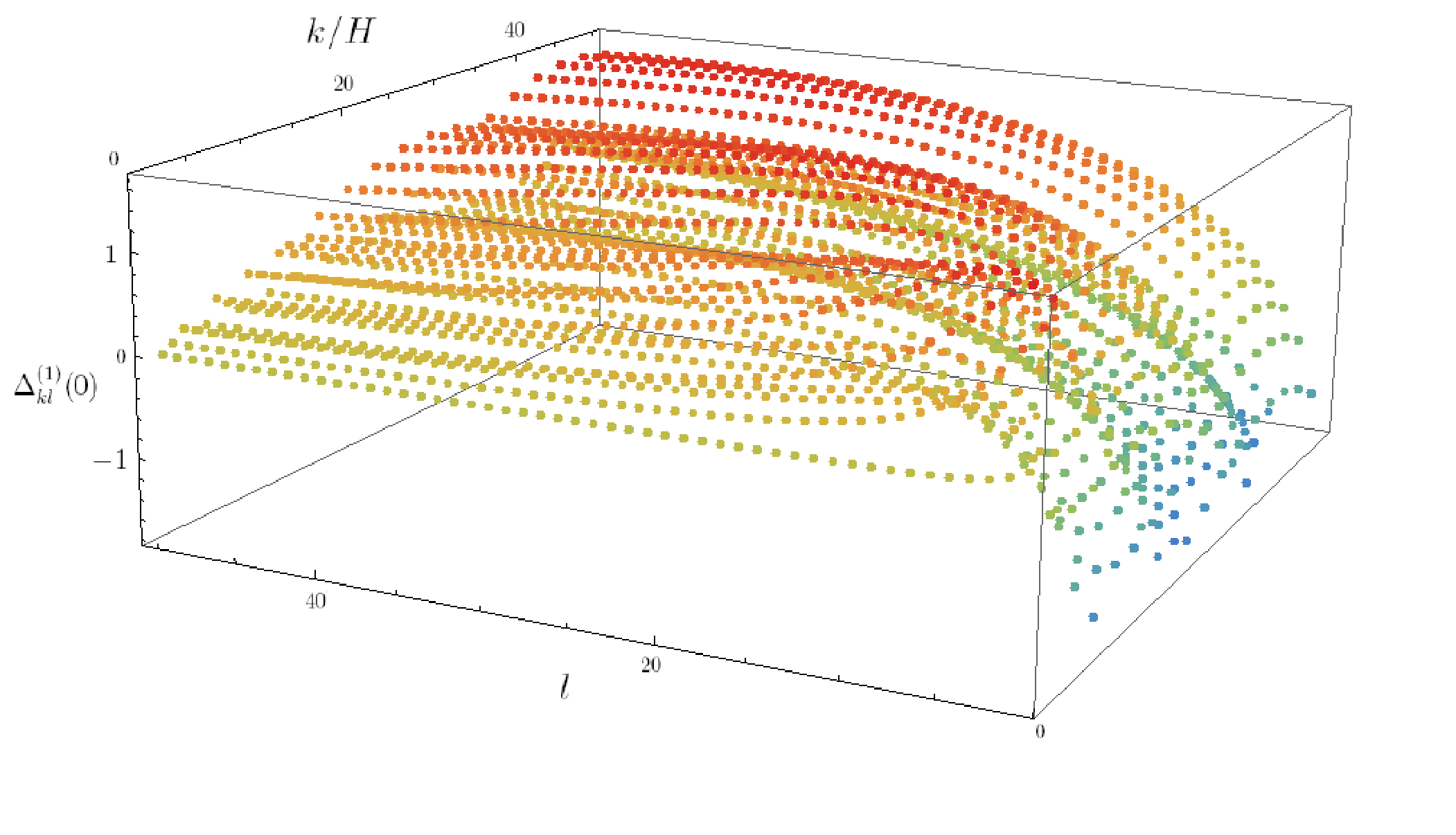}
\caption{Three-dimensional plot of the asymptotic values $\Delta^{(1)}_{kl}(0)$
in Eq.~(\ref{deltalk0}) against $l$ and $k/H$ for $1<l<50$ and $1<k/H<50$.}
\label{fig2}
\end{figure}

In Fig.~\ref{fig1}, we show the time evolution of the first-order
contribution to the scalar fluctuations from Eq.~(\ref{plk}).
Actually, it is more convenient to plot the normalized power spectrum
$\Delta^{(1)}_{kl}(\tau)$ in Eq.~(\ref{deltalk})
against the e-folding number $Ht$. We have chosen the
angular momentum $l=2,22,40$ as examples. For each $l$,
different $k$ modes are shown. From all the plots, we can see
that for a k-mode $\Delta^{(1)}_{kl}(\tau)$ oscillates when the mode is still
sub-horizon. Once the mode crosses out the horizon, $\Delta^{(1)}_{kl}(\tau)$
stop oscillating and gradually approaches a constant
value. This behavior can be easily explained by Eq.~(\ref{1os}),
where the source term dies off as $\varphi_{kl} ^{(0)}$ goes
super-horizon and then gets frozen. In Fig.~\ref{fig2}, we plot
the asymptotic values $\Delta^{(1)}_{kl}(0)$ in Eq.~(\ref{deltalk0}) for
$1<l<50$ and $1<k/H<50$. The figure shows that $\Delta^{(1)}_{kl}(0)$ is suppressed
in low-$l$ and low-$k$ regions. This can be explained considering that in
this limit one is considering fluctuations on large scales, where the effects of the black
hole should be negligible.
Also, the magnitude of $|\Delta^{(1)}_{kl}(0)|$ is
of order one, so the first-order contribution, compared to the
zero-order de Sitter power spectrum, is roughly downsized by the
expansion parameter $\epsilon$.

For the quantum case, we plot the asymptotic values $\Delta^{(1)}_{kl}(r,0)$
in Eq.~(\ref{delta1klr0}) for $Hr=0.01,\,0.1$ against $l$ and $k/H$
for $1<l<50$ and $1<k/H<50$. Note that if inflation lasts for about $60$ e-folds,
$r=H^{-1}$ will be about the size of the present Universe and $Hr<1$ corresponds
to sub-horizon length scales.
As shown in Fig.~\ref{fig3}, the general trend is similar to the classical case
in Fig.~\ref{fig2}, except that there are some differences both at low $l$ and low $k$.
This is indeed an explicit example that quantum fluctuations generated during
inflation behave like classical waves. We have also calculated
$\Delta^{(1)}_{kl}(r,0)$ for larger values of $Hr$,
which becomes fluctuating but in general the value of the amplitude is getting smaller.
It is expected because the farther the black hole is, the lesser is its effect to
the perturbation.

\section{Conclusions}
\label{conclusion}

We have presented a perturbation method to compute the effect of the
presence of a black hole in the de Sitter space to the quantum
fluctuations of a free massless scalar field. The method is valid as
long as the expansion parameter $\epsilon\equiv GMH\ll 1$, i.e.,
the size of the black hole event horizon is smaller than that of the de Sitter
cosmological horizon. The calculation can be easily applied to
a vector field or a gravitational wave. Here the first-order contribution is
computed and the results are given in the assumption that the black hole is
located at the origin of the coordinates. Higher-order corrections can be
worked out perturbatively though complicated. It would be interesting
to consider the effect due to a distribution of black holes in the de Sitter
space. In fact, the perturbation can be in a form of cosmological defects such
as monopoles, cosmic strings, or domain walls.

Let us briefly discuss some cosmological implications of the results that
we have obtained in this work. Assume that inflation lasts for
about $60$ e-folds. Then, the wavelength of the Fourier mode with $k/H=1$
is about the size of the present Universe.
If the location of the black hole that exists during inflation is near the Earth,
the suppressed power of the first-order correction to the de Sitter
inflaton fluctuations in low $l$ and low $k$ regions may result
in a blue-tilted density power spectrum on large angular scales.
This in turn gives rise to a suppression of the large-scale
cosmic microwave background anisotropy that may be relevant to the
observed low quadrupole in the WMAP cosmic microwave background
anisotropy data~\cite{wmap5b}. A detailed calculation of the
effect to the cosmic microwave background anisotropy is underway,
including the case that the black hole locates somewhere else in
the Universe.

\begin{figure}[htp]
\centering
\includegraphics[width=0.68\textwidth]{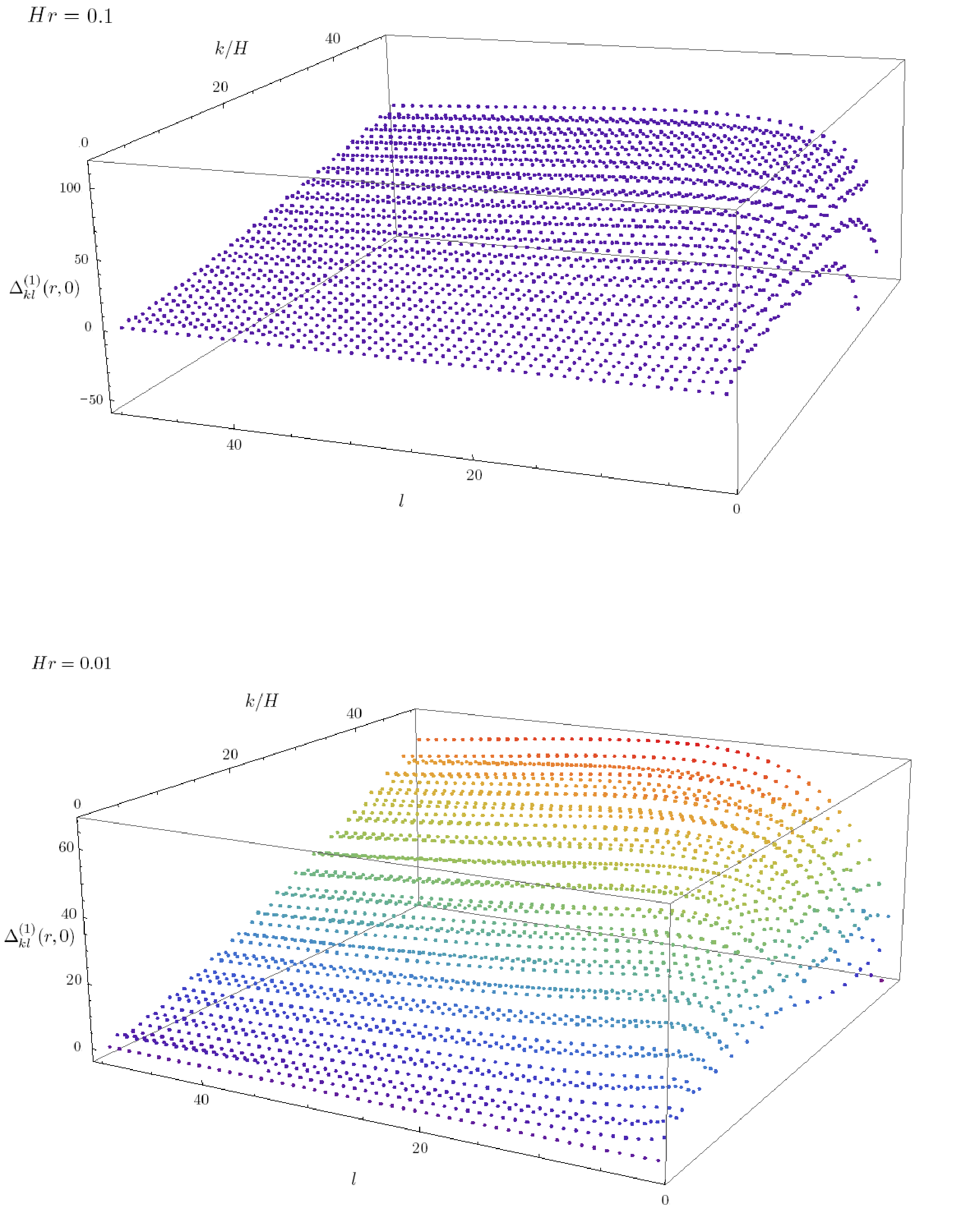}
\caption{Three-dimensional plots of the asymptotic values $\Delta^{(1)}_{kl}(r,0)$
in Eq.~(\ref{delta1klr0}) for $Hr=0.01$ (lower) and $Hr=0.1$ (upper) against $l$ and $k/H$ for $1<l<50$ and $1<k/H<50$.}
\label{fig3}
\end{figure}

\begin{figure}[htp]
\centering
\includegraphics[width=0.68\textwidth]{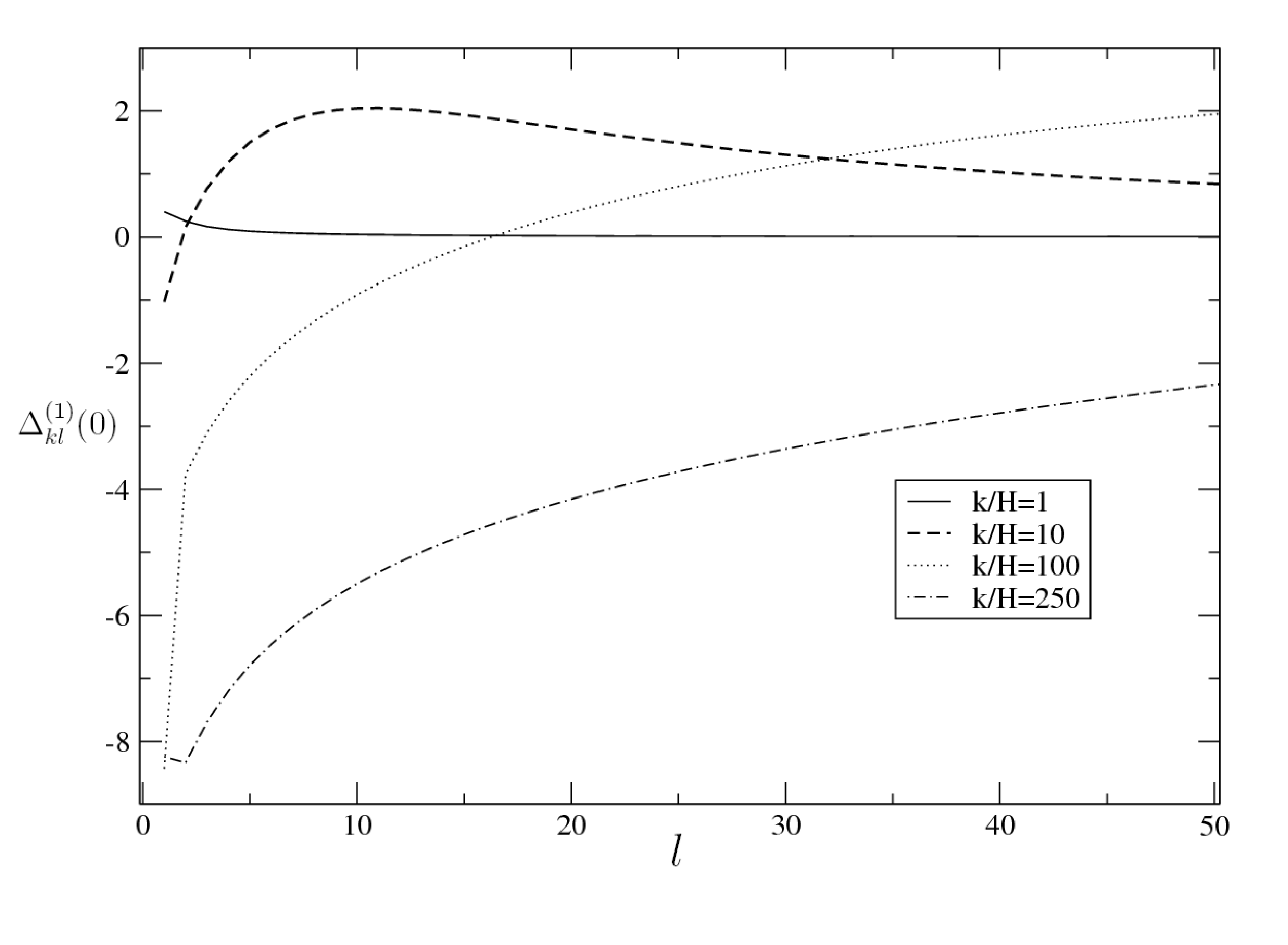}
\caption{Asymptotic values $\Delta^{(1)}_{kl}(0)$ against $l$ for
$k/H=1$, $10$, $100$, and $250$.} \label{fig4}
\end{figure}

If inflation lasts for a longer time, then the wavelength of the
Fourier mode that corresponds to the size of the present Universe
will be given by a larger value of $k/H$. In Fig.~\ref{fig4}, we
plot the asymptotic values $\Delta^{(1)}_{kl}(0)$ against $l$ for
$k/H=1$, $10$, $100$, and $250$, which correspond to the inflation
with $60$, $62.3$, $64.6$, and $65.5$ e-folds, respectively. As
expected, the longer is the inflation duration the lesser
pronounced are the effects of the black hole to large-scale or
low-$l$ observations. This can also be seen in the quantum case
with $\Delta^{(1)}_{kl}(r,0)$ in Fig.~\ref{fig3}.

It is worth noting that the present work gives a
realization of the general discussions in Ref.~\cite{carroll}
about the potentially observable effects of a small violation of
translational invariance during inflation, as
characterized by the presence of a preferred point, line, or plane.
This violation may induce derivations from pure statistical isotropy of
cosmological perturbations, thus leaving anomalous imprints on the
cosmic microwave background anisotropy~\cite{carroll}.

\begin{acknowledgments}
We thank Lau-Loi So for his contributions at the initial stage of
the paper and the anonymous referee for his/her comments.
This work was supported in part by the National Science
Council, Taiwan, ROC under the Grants NSC 96-2112-M-032-006-MY3
(HTC), NSC 95-2112-M-001-052-MY3 (KWN), NSC 97-2112-M-003-004-MY3
(ICW), and the National Center for Theoretical Sciences, Taiwan,
ROC.
\end{acknowledgments}

\end{document}